# The water clock of the Bronze Age (Northern Black Sea Coast)

**Larisa N. Vodolazhskaya[1], Anatoliy N. Usachuk[2], Mikhail Yu. Nevsky[3]**

[1] Southern Federal University (SFU), Rostov-on-Don, Russian Federation;
E-mails: larisavodol@aaatec.org, larisavodol@yahoo.com
[2] Donetsk Regional Museum, Donetsk, Ukraine; E-Mail: doold@mail.ru
[3] Southern Federal University (SFU), Rostov-on-Don, Russian Federation; E-mail: munevsky@sfedu.ru

**Abstract**

In the article presents the results of the multidisciplinary study conducted with the help of archaeological, physical and astronomical methods. The aim of the study was to analyze and interpret marks and drawings applied to the surface of the vessel of the Bronze Age (Srubna culture) found near the Staropetrovsky village in the northeast of the Donetsk region (the Central Donbass). The carried out calculations and measurements possible to prove that Staropetrovsky vessel is the most ancient water clock, discovered on the territory of Europe, and have approximately the same age as the oldest known ancient Egyptian water clock. Such vessels - water clocks were needed for Srubna population to rituals related to the determined time of day and to mark sundial, which had recently been discovered in the Northern Black Sea Coast. Staropetrovsky vessel has marks of heliacal rising of Sirius and it is an ancient astronomical auxiliary instrument for determining the time at night.

**Introduction**

In 1985, near the village of Staropetrovsky (neighborhood Yenakiyevo, Donetsk region) a clay pot, owned Srubna culture and dating XV-XIV centuries BC, was found in a ruined barrow [1-3].

Its uniqueness lies in the combination groups of signs printed on the outer and inner surface of the vessel. Especially rare is the label on the inside of the vessel, which is a vertical row of nail marks (Fig. 1).

At the time, preparing materials of excavated kurgans of the Central Donbas for publication, the authors placed a description of the Staropetrovsky vessel in the section "The findings from the destroyed kurgans" [4]. At that time, the vessel was kept in the archaeological collection of the head of Yenakiyevo archaeological expedition V. F. Klimenko [5]. In 2002, ending almost 30 years of work [6] V. F. Klimenko handed over all materials to the funds of the Donetsk Regional Museum. Staropetrovsky vessel (museum index: DOKM a 3161, kp 112219) was accepted for storage in a museum together with a numerous funerary ware.

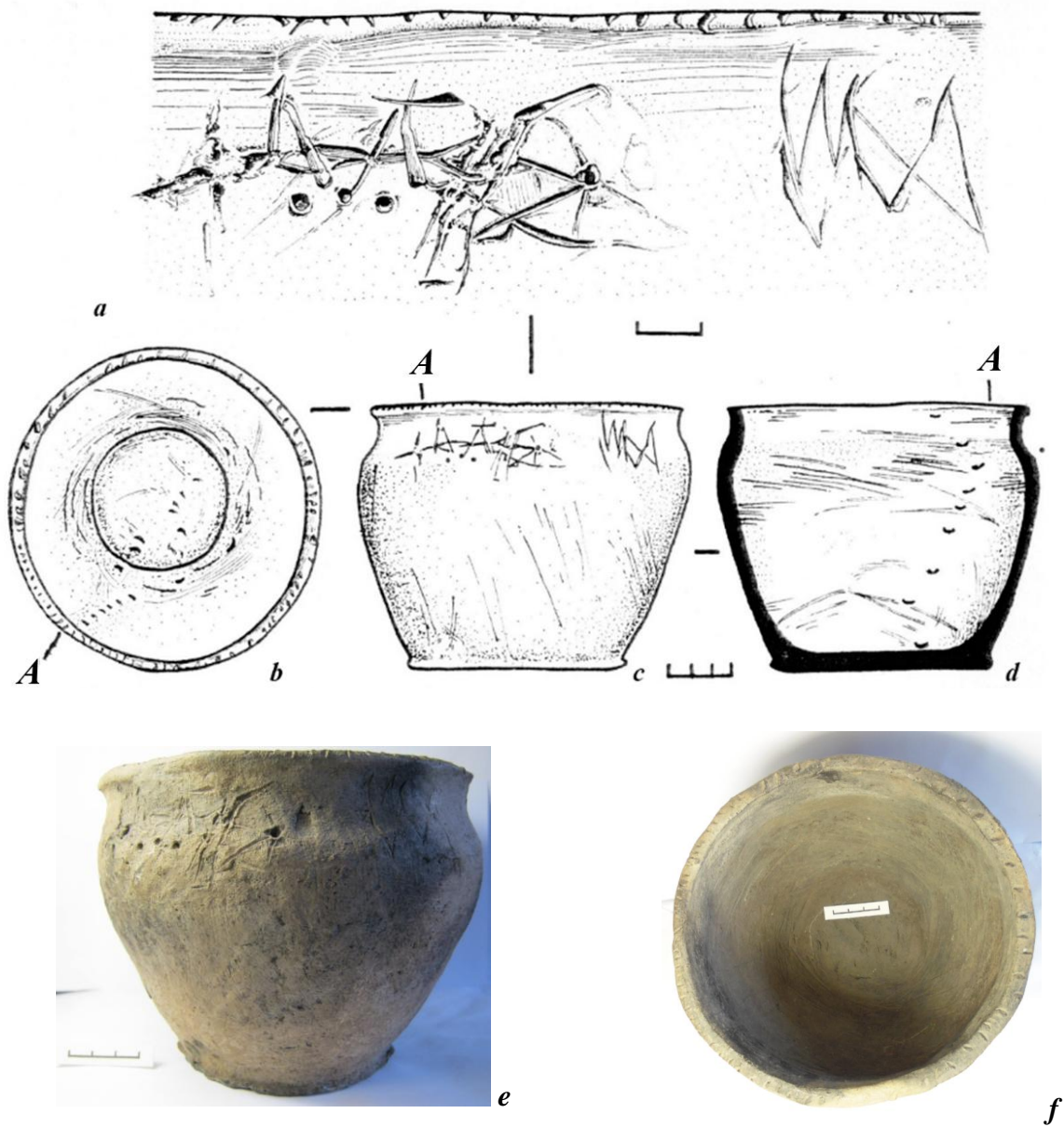

**Figure 1.** Vil. Staropetrovsky, ruined barrow, vessel with the marks on the inner surface: *a* - drawing composition on the outer side of the vessel, *b* - top view (drawing), *c* - a side view (drawing), d - marks on the inner side surface (drawing); *A* - place labeling on the inner surface relative the compositions on the outside of the vessel (drawing: V.B. Punkovsky, A.N. Usachuk, 1993), *e* - photo of the outer side surface of the vessel (photo by A.N. Usachuk, 2014), *f* - photo inner surface of the vessel (photo by A.N. Usachuk, 2014).

**Archaeological description of the vessel**

The carinated vessel was made in the following mode: first a cylindrical body 8.8 to 9.4 cm in height was formed, and then a 3 cm wide clay coil was applied. The inflection formed with the ridge was slightly corrected and leveled, leaving shallow finger impressions. The rim of the vessel is slightly everted, while the base meets the body at a concave angle. Near the base are horizontal impressions which were left by the fingers while attaching the base to the body. The interior surface of the left side of the base is very irregular and there remain long traces of the evening and smoothing of the raw clay with fingers. The exterior surface of the vessel is also irregular, with areas of fine horizontal and diagonal impressions made by a wooden instrument.

These impressions lie on top of the fine combing which is almost covered by subsequent smoothing. The comb stamp was used for a more even distribution of clay on the vessel's surface prior to smoothing [7].

Under the rim is a band of horizontal finger impressions made while forming the vessel. It should be mentioned that the part of the surface with a group of incised marks was deliberately made smoother. On the opposite side of the vessel, the surface area between the rim and the ridge was not regularized, leaving coarse horizontal-diagonal smoothing walls.

The interior vessel surface has deep traces of horizontal and diagonal smoothing by a wooden chip and sometimes, under these traces, are weak horizontal finger impressions. Once the vessel was smoothed, several diagonal incised lines were made. One incision has a width of 0.1-0.2 cm and travels abruptly from the vessel bevel to its base, erasing the traces of smoothing. This incision was perhaps formed in a single movement. Because this incision, made with a thin (probably wooden) tool, is clear, deep and narrow at the base and becomes shallower, wider and has an irregular section at the rim, it can be concluded that the application of the track went from the base to the mouth of the vessel.

Inside the vessel, just over 5 cm from this line, two thin diagonal marks parallel to each other were incised on the wet clay with a pointed tool (of unclear material: metal needle ?, very thin metal blade?). In contrast to the wide lines these thin marks start under the rim, on the ridge inside the vessel, and nearly reach the bottom. These marks were made with one movement each, applied in the opposite direction to the moves which were used to make the wider lines: from the inside of the ridge down to the bottom of the vessel. From a technological point of view, diagonal wide and narrow tracks are unnecessary in the manufacturing of the vessel[1].

Could the described manipulation of an almost finished product be associated with some ritual practices before drying and firing?[2] It should be mentioned again that the wide and narrow tracks were made in opposite directions. Similar idea lies in the pottery spatula from the Timber Grave settlement Ilichevka, located in the Don basin (Donetsk region, Ukraine) [8]. In this case, on the surface of a tool made from a fragment of the cattle rib, an image of two birds that are opposed to each other (legs facing in opposite directions) was engraved. The opposition of these birds was re-enforced by the fact that the images were cut using two different tools (flint and bronze) [9]. The Ilichevka birds on the potter's spatula did not go unnoticed [10, 11]. Perhaps the idea of the opposing images on those potter's tools and the opposing diagonal incised marks inside the vessel are similar.

The exterior surface under the rim of the vessel found in Staropetrovsky was covered with incised marks (Fig. 1). There are two groups of the marks: one having the width of 8.2 cm and the other 3 cm. Roughly incised polyline ("zigzag") lines form the long composition. On the left is located a roughly incised saltire, on the right is a round composition, which resembles a wheel with spokes. Over the long composition on the right side a short horizontal line was made. This saltire was not depicted by two lines but by four scratches from the center, displacing a small portion of clay which was sloppily smoothed by a finger. Under the polyline, three impressions are located. The diameter of the side impressions measures 0.2 cm, the central one - 0.15 cm.

---

[1] Such an operation, though on the exterior surface of the molded dishes, has been recorded in the manufacture of cookware in the Timber Grave settlement Shirokaya Balka-II in Northeast Azov Sea [12].

[2] Pay attention to the fact that the potters were aware that a turning point in cookware manufacturing falls at the beginning of the firing process [13]. Various rites are performed at this time: they place next to the vessel in the kiln food offerings, cross-mark the kilns with a candles, making sure that at this time no strangers were near the hearth, especially women, sometimes bypass the kiln for three times and make a fire with a willow sprig [14].

Their depth is 0.3 cm. The central impression closes the polyline line. The "wheel" is a semicircle which was incised to the right of the polyline in a rough manner (in three stages). This semicircle was incised from upside down, and near the second line (in order of making) two irregular thin lines are located parallel to both each other and the second line of the semicircle. In the middle of "the wheel", a circular impression 0.3 cm in diameter and 0.45 cm deep was placed. From these impressions five "spokes" are depicted. Judging by the imposition of incised lines, the composition was applied from right to the left: first the impression was made, then the "spokes", and then a semicircle. Initially, the first, second, fourth and fifth (very weak) "spokes" were outlined (counting from top to bottom from left to right). The third "spoke" was depicted after the semicircle.

After the semicircle, two roughly sloping lines were depicted, then a short horizontal line, and then three impressions were made. The impressions were followed by a polyline, and then the cross. The lines were incised with the edge of a wooden tool, obviously, the same which was used for making the diagonal mark on the interior surface of the vessel. Located 1.8 cm to the right of the first group, the second group was placed, presenting as a rough polyline. The lines of this polyline were narrow. The lines of the second group were made with a finer tool than of the first one.

The bevel was decorated with nail imprints, placed more frequently in the area below the rim where the two described groups of lines are located. On the exterior surface of the vessel, in front of the cross from the first group, a vertical irregular line of nail imprints passes, beginning in the ridge of the vessel and the going down to the bottom. A total of eight nail prints were placed on the body of the vessel, and eight prints on the bottom. Inside the vessel, 2.6 cm to the left of the 7th from the bottom nail print, two more nail prints are located. They are more shallow and have a less clear outline than almost all the prints in the line (Fig. 1).It seems that they are not included in the vertical line of nail impressions[3]. Two thin scratches going from the inside ridge to the bottom of the vessel are opposite of the nail prints.

Impressions №№ 2-8 were made with the same nail. The length of the prints is 0.55 cm. At the bottom of the vessel the prints measure more: 0.6-0.65 cm. Their thickness is 0.1 cm and sometimes a little more. Given the thickness of the prints, they belonged to an adult [15] who, when applying, likely did not use the entire nail plate, but only its regrown region.

The vessel has an irregular shape. Its height is 11.6-11.8 cm and 12.0-12.2 cm. The rim diameter is 14.3-14.6 cm, the ridge diameter is 15.3-15.6 cm, and base diameter- 9.4-9.8 cm (the base edge in some places a little lost).

The Staropetrovsky vessel is yellow-gray with dark gray (sometimes almost black) spots. Judging from these spots, open firing was used with reducing conditions [16-18]. The prevailing light gray color of the interior surface suggests that the vessel was standing on its base during the firing process and was filled with ash [19]. The fabric of the sherd cross-section is of a dark gray color without abrupt layers, reflecting the long period when the vessel sat in the firing pit [20, 21]. In general, the firing of the vessel was good enough that the vessel walls sound clear and almost shrill. The quality of the vessel was tested during experiments by repeatededly filling it with water (see. Below). Most likely, the vessel of Staropetrovsky was burned in a fire but not in a bonfire as the firing device was protected from the wind. It could be a pit [22, 23], or ash-filled pit [24]. Such a technology enables to smooth negative aspects of open firing [25] and was widely known in Timber Grave context [26, 27].

---

[3] The irregularity of these three impressions and lack of connection with the vertical line of nail impressions led to the fact that, in the description of the vessel made in 1994, this detail was missed.

At the moment, a unique vessel is located in Donetsk regional museum. The vessel was not damaged during the shelling of the museum in August 2014[4].

**Calculations of volume of Staropetrovsky vessel**

Discovered in 1985 near-by the settlement of Staropetrovsky Srubna vessel has marks on an internal side, reminding marking of measuring vessel. To prove that the discovery is really, measuring vessel, we conducted interdisciplinary studies with the help of natural science methods, which have been particularly widely used in archeology in recent years [28-35].

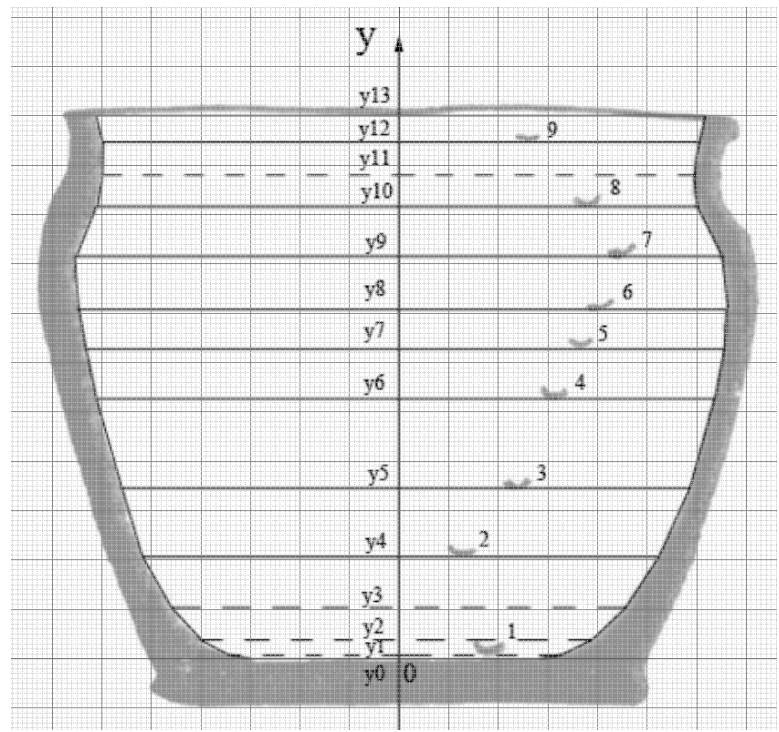

**Figure 2.** Vil.Staropetrovsky, ruined barrow, vessel with the marks on the inner surface: Y is the vertical co-ordinate axis directed from a bottom to the halo of vessel, $y_i$ - coordinates of marks and auxiliary layers on the axis of Y. Next to marks their numbers are filled in.

Originally, on the basis of drawing published before (fig. 1), we conducted the calculations of volumes, corresponding to the marks of vessel. A vessel was mentally divided into eight horizontal layers between nine marks. From the steep bends of lateral walls, for more exact calculation of volume of vessel, layer between the first and second marks was additionally divided into three layers, and layer between eighth and ninth marks - on two layers. Space between a bottom and the first mark, and also by a ninth mark and overhead edge of vessel, also examined by us, as separate layers. Thus, a vessel was mentally broken by us on thirteen layers (N=13).

The side of vessel we approximated a polyline corresponding to dividing of vessel by layers (fig. 2). Every layer was presented by us, as the truncated sloping abrupt the volume of that settled accounts on a formula 1:

---

[4] http://www.youtube.com/watch?v=FKkCIKYM2dc

$$V_i = \frac{1}{12} \cdot \pi \cdot h_i \cdot (d_{i-1}^2 + d_{i-1} \cdot d_i + d_i^2),$$
$$\text{где } h_i = y_i - y_{i-1},$$
$$i \in [1; N] \qquad (1)$$

$V_i$ – calculated volume of $i$ cone, $i$ – number of cone, $y_i$ - coordinate of mark of vessel or auxiliary layer on the axis of Y (for the bottom $i=0$), $h_i$ – height of $i$ cone, $d_i$ – diameter of founding of $i$ cone, $N$ – common amount of layers. In our case $N=13$.

The results of our calculations are presented in the table 1.

**Table 1.** Calculated volumes of layers - the truncated cones; $i$ – number of cone, $y_i$ - coordinate of mark of vessel or auxiliary layer on the axis of Y (for the bottom $i=0$), $h_i$ – height of $i$ cone, $d_i$ – diameter of founding of $i$ cone, $V_i$ – calculated volume of $i$ cone.

| $i$ | $y_i$, (cm) | $h_i$, (cm) | $d_i$, (cm) | $V_i$, (cm$^3$) |
|---|---|---|---|---|
| 0 | 0.0 | 0.0 | 5.8 | - |
| 1 | 0.1 | 0.1 | 6.7 | 3.1 |
| 2 | 0.4 | 0.3 | 7.9 | 12.6 |
| 3 | 1.1 | 0.7 | 9.2 | 40.0 |
| 4 | 2.1 | 1.0 | 10.5 | 75.7 |
| 5 | 3.5 | 1.4 | 11.5 | 132.9 |
| 6 | 5.3 | 1.8 | 12.5 | 203.6 |
| 7 | 6.3 | 1.0 | 12.9 | 126.5 |
| 8 | 7.2 | 0.9 | 13.1 | 119.5 |
| 9 | 8.2 | 1.0 | 13.1 | 134.3 |
| 10 | 9.2 | 1.0 | 12.0 | 123.7 |
| 11 | 9.8 | 0.6 | 11.9 | 67.6 |
| 12 | 10.5 | 0.7 | 12.0 | 78.8 |
| 13 | 11.0 | 0.5 | 12.2 | 57.9 |

Calculated by us on drawing the total volume of vessel is equal to the sum of volumes of all layers of $V_{all}$=1176.2 cm$^3$.

The resulting volumes of layers between marks are presented in the table 2.

**Table 2.** Calculated resulting volumes of layers - the truncated cones between marks; $m$ - number of mark, $j$ - number of layer between marks, $V_j$ - volume of $j$ layer.

| $m$ | $j$ | $V_j$, см$^3$ |
|---|---|---|
| 1 | - | - |
| 2 | 1 | 128.3 |
| 3 | 2 | 132.9 |
| 4 | 3 | 203.6 |
| 5 | 4 | 126.5 |
| 6 | 5 | 119.5 |
| 7 | 6 | 134.3 |
| 8 | 7 | 123.7 |
| 9 | 8 | 146.4 |

The measured maxheight of Staropetrovsky vessel is equal 12.2 cm [36].

The volume of vessel between extreme (1th and 9th) marks settled accounts by us on formula 2, mean volume of layer between marks - on a formula 3, and standard deviation settled accounts on the formula 4 [37]:

$$V_{marks} = \sum_{j=1}^{n} V_j \qquad (2)$$

$$V_{ar\_m} = \sum_{j=1}^{n} \frac{V_j}{n} \qquad (3)$$

$$\sigma = \pm \sqrt{\frac{1}{n}\sum_{j=1}^{n}(V_{ar\_m} - V_j)^2} \qquad (4)$$

where $V_{marks}$ - volume of vessel between extreme (1th and 9th) marks, $j$ - number of layer between marks, $n$ - common amount of layers between marks (n=8), $V_j$ – volume of $j$ layer, $V_{ar\_m}$ - mean volume (arithmetical mean) of layer between marks, $\sigma$ - standard deviation.

Expected on a formula 2 volume of vessel between the extreme (1th and 9th) marks of $V_{marks}$=1115.3 cm$^3$. Expected on a formula 3 mean volume of layer between the marks of $V_{ar\_m}$=139.4 cm$^3$. Expected on a formula 4 standard deviation $\sigma$=±25.4 cm$^3$. Thus, the volume of every layer of vessel between two nearby marks appeared equal 139.4±25.4 cm$^3$. The volumes of almost all layers got in this range. Only the volume of 3th layer on 38.8 cm$^3$ (23.5%) exceeded the high bound of the indicated range. For finding out of reason of such rejection realization of the direct measuring of volume of vessel was required by water. If reason consisted in inaccuracy of drawing, then such measuring fully would confirm our hypothesis that a Staropetrovsky vessel could be used as clepsydra.

Thus, the calculations conducted by us on the basis of the primary drawing of Staropetrovsky vessel, in the first approaching, confirmed a hypothesis that a Staropetrovsky vessel could be the clepsydra of story type for measuring of temporal intervals of equal duration, as marks marked volumes approximately the same size. However, for a final conclusion, measuring of volume of vessel was required by water.

**Measuring of volume of Staropetrovsky vessel**

For verification of results of calculations and clarification of volume, A.N. Usachuk conducted measuring of volume of every layer of vessel between marks by means of water. In an order to obtain maximal exactness of measuring, chemical laboratory measure glasses were used: on 200 ml (№ 00159) and on 25 ml (without №) of PAO "Steklopribor".

In the process of the repeated study of marks it was educed on the internal side of Staropetrovsky vessel, that 1-th mark on drawing behaves to the group of marks on the bottom of vessel, as is on continuation of line along that most marks of bottom are situated. However, approximately on the same height, but on continuation of line along that most lateral marks are situated, there is another mark not marked on the primary drawing. This mark and was numbered by us in measuring, as a 1-th mark (fig. 3).

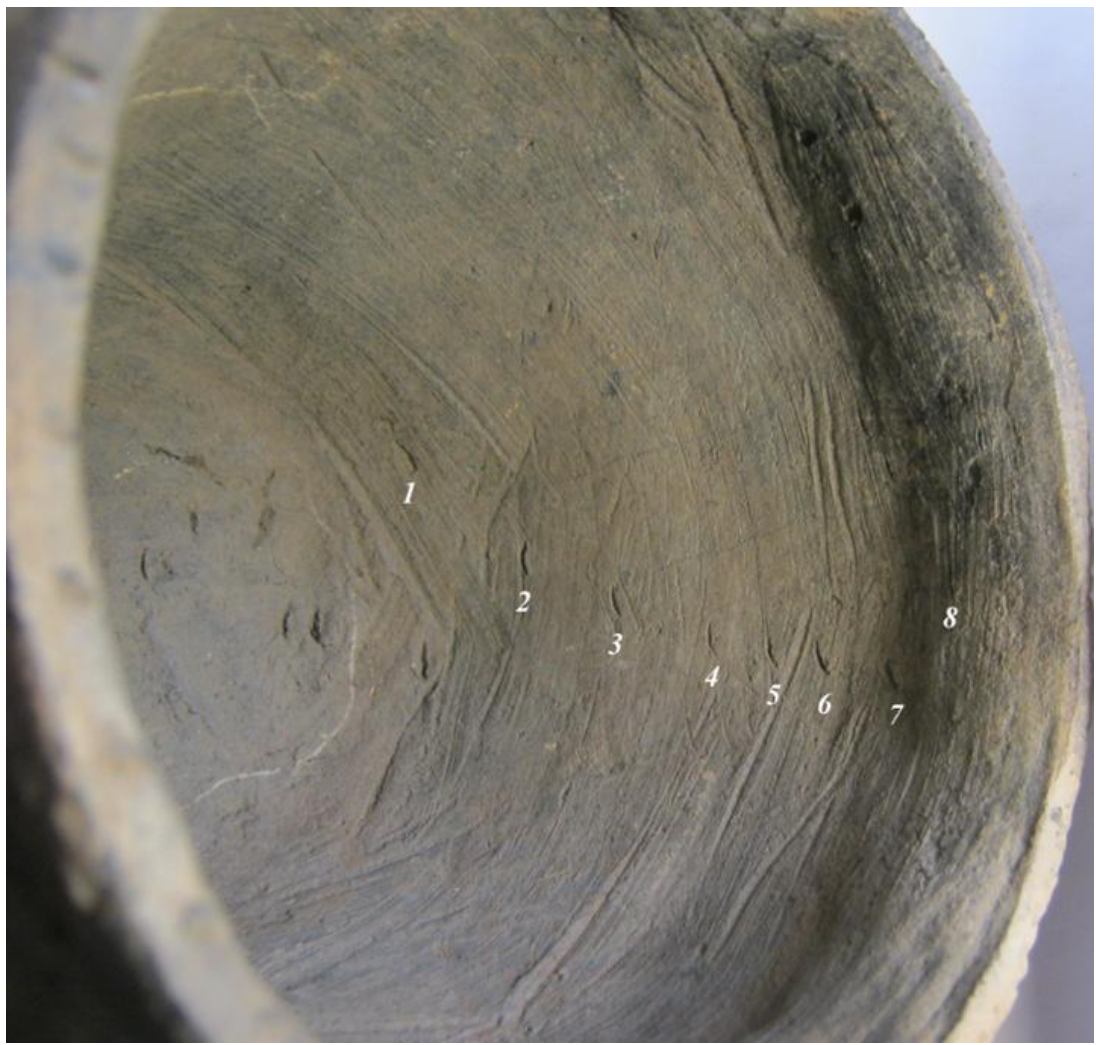

**Figure 3.** Vil. Staropetrovsky, ruined barrow, vessel with the marks on the inner surface, photo of internal surface of vessel. On a photo next to marks on the side of their number in measuring.

It was also discovered that the 9th mark of drawing by sight notedly differs from other marks and actually is a casual scratch, therefore by us it was not examined in measuring. And, as a result, measuring of volume between 8th and 9th marks not produced.

At comparison of location of marks with a picture on the external side of vessel it was educed, that mark 7 is situated approximately at the same level, what underbody of rim of "wheel", and mark 8 - at the level of center of wheel. I.e. the elements of "wheel" could perform the duty of marks on exteriority of vessel. Supposing that overhead part of rim of "wheel" also could act part mark, we made an effort take into account and it in the further measuring water, designating, as a mark of *T*.

Every measuring of vessel water was produced in six repetitions. Water was poured by means of measure glasses (determining a volume to the same) to each of marks and for marks on an internal side measured the height of water level from a bottom. The results of measuring are presented in a table 3 and 4.

**Table 3.** Results of measuring of volume of vessel by water; $V_{m\_1},...,V_{m\_8}$ is the measured volume of vessel from a bottom to each of eight marks, $V_{m\_T}$ of $T$ is the measured volume of vessel to the mark of $T$, $V_{m\_gen}$ - the general measured volume of vessel.

| measuring number | $V_{m\_1}$, (ml) | $V_{m\_2}$, (ml) | $V_{m\_3}$, (ml) | $V_{m\_4}$, (ml) | $V_{m\_5}$, (ml) | $V_{m\_6}$, (ml) | $V_{m\_7}$, (ml) | $V_{m\_8}$, (ml) | $V_{m\_T}$, (ml) | $V_{m\_gen}$, (ml) |
|---|---|---|---|---|---|---|---|---|---|---|
| 1 | 36.5 | 140.0 | 259.5 | 418.0 | 539.0 | 686.5 | 848.0 | 999.0 | 1138.5 | 1187.5 |
| 2 | 36.0 | 143.0 | 256.0 | 414.6 | 532.5 | 684.0 | 846.0 | 988.8 | 1138.3 | 1180.5 |
| 3 | 37.0 | 137.5 | 260.5 | 410.5 | 528.2 | 679.5 | 842.5 | 983.0 | 1137.5 | 1181.0 |
| 4 | 37.6 | 137.0 | 264.9 | 416.3 | 525.0 | 690.5 | 842.0 | 988.0 | 1134.5 | 1180.0 |
| 5 | 36.4 | 143.3 | 257.8 | 411.5 | 523.5 | 686.0 | 846.5 | 984.0 | 1128.5 | 1186.5 |
| 6 | 37.0 | 140.0 | 252.0 | 414.5 | 523.0 | 682.5 | 842.5 | 996.0 | 1133.8 | 1171.0 |

**Table 4.** Results of measuring of height of location of marks in relation to a bottom; $y_{m\_1},...,y_{m\_8}$ – measured height of each of eight marks.

| measuring number | $y_{m\_1}$, (cm) | $y_{m\_2}$, (cm) | $y_{m\_3}$, (cm) | $y_{m\_4}$, (cm) | $y_{m\_5}$, (cm) | $y_{m\_6}$, (cm) | $y_{m\_7}$, (cm) | $y_{m\_8}$, (cm) |
|---|---|---|---|---|---|---|---|---|
| 1 | 0.8 | 2.3 | 3.7 | 5.1 | 6.1 | 6.9 | 8.2 | 9.4 |
| 2 | 0.8 | 2.2 | 3.7 | 5.1 | 6.1 | 6.9 | 8.2 | 9.4 |
| 3 | 0.7 | 2.3 | 3.8 | 5.1 | 6.0 | 6.9 | 8.2 | 9.4 |
| 4 | 0.8 | 2.3 | 3.9 | 5.1 | 6.0 | 6.9 | 8.2 | 9.4 |
| 5 | 0.7 | 2.4 | 3.8 | 5.1 | 6.0 | 6.9 | 8.2 | 9.5 |
| 6 | 0.8 | 2.4 | 3.8 | 5.1 | 6.1 | 6.9 | 8.2 | 9.5 |

Then, the got volumes and heights were counted for every layer on formulas 5 and 6:

$$V_{l\_j} = V_{m\_j} - V_{m\_(j-1)} \tag{5}$$

$$h_{l\_j} = y_{m\_j} - y_{m\_(j-1)}, \tag{6}$$

where $j \in [1; k]$

where $j$ - number of layer, $n$ - amount of layers ($k=7$), $V_{l\_j}$ - measured volume of layer $j$, $V_{m\_j}$ - measured volume of vessel from a bottom to the mark $m=j+1$, $h_{l\_j}$ is the measured height of layer $j$, $V_{m\_j}$ - measured height of vessel from a bottom to the mark $m=j+1$.

The results of such count are presented in a table 5 and 6.

**Table 5.** Volume of layers of vessel between marks; $V_{l\_0}$ - volume of layer of vessel from a bottom to the mark *1*, $V_{l\_0}$,., $V_{l\_7}$ - height of layers between nearby eight marks, $V_{l\_T}$ - volume of layer of vessel from a mark 8 to the mark *T*, $V_{l\_h}$ - volume of layer of vessel from the mark *T* to the edge of corolla.

| measuring number | $V_{l\_0}$, (ml) | $V_{l\_1}$, (ml) | $V_{l\_2}$, (ml) | $V_{l\_3}$, (ml) | $V_{l\_4}$, (ml) | $V_{l\_5}$, (ml) | $V_{l\_6}$, (ml) | $V_{l\_7}$, (ml) | $V_{l\_T}$, (ml) | $V_{l\_h}$, (ml) |
|---|---|---|---|---|---|---|---|---|---|---|
| 1 | 36.5 | 103.5 | 119.5 | 158.5 | 121.0 | 147.5 | 161.5 | 151.0 | 139.5 | 49.0 |
| 2 | 36.0 | 107.0 | 113.0 | 158.6 | 117.9 | 151.5 | 162.0 | 142.8 | 149.5 | 42.2 |
| 3 | 37.0 | 100.5 | 123.0 | 150.0 | 117.7 | 151.3 | 163.0 | 140.5 | 154.5 | 43.5 |
| 4 | 37.6 | 99.4 | 127.9 | 151.4 | 108.7 | 165.5 | 151.5 | 146.0 | 146.5 | 45.5 |
| 5 | 36.4 | 106.9 | 114.5 | 153.7 | 112.0 | 162.5 | 160.5 | 137.5 | 144.5 | 58.0 |
| 6 | 37.0 | 103.0 | 112.0 | 162.5 | 108.5 | 159.5 | 160.0 | 153.5 | 137.8 | 37.2 |

**Table 6.** Height of layers between marks; $h_{l\_0}$ - height of layer from a bottom to the mark 1, $h_{l\_1},..., h_{l\_7}$ – height of layers between nearby marks.

| measuring number | $h_{l\_0}$, (cm) | $h_{l\_1}$, (cm) | $h_{l\_2}$, (cm) | $h_{l\_3}$, (cm) | $h_{l\_4}$, (cm) | $h_{l\_5}$, (cm) | $h_{l\_6}$, (cm) | $h_{l\_7}$, (cm) |
|---|---|---|---|---|---|---|---|---|
| 1 | 0.8 | 1.5 | 1.4 | 1.4 | 1.0 | 0.8 | 1.3 | 1.2 |
| 2 | 0.8 | 1.5 | 1.5 | 1.4 | 1.0 | 0.8 | 1.3 | 1.2 |
| 3 | 0.7 | 1.6 | 1.6 | 1.3 | 0.9 | 0.9 | 1.3 | 1.2 |
| 4 | 0.8 | 1.5 | 1.6 | 1.3 | 0.9 | 0.9 | 1.3 | 1.2 |
| 5 | 0.7 | 1.7 | 1.5 | 1.3 | 1.0 | 0.9 | 1.3 | 1.3 |
| 6 | 0.8 | 1.7 | 1.4 | 1.4 | 1.0 | 0.9 | 1.3 | 1.3 |

Mean height of layer between marks and mean volume of layer between marks were calculated for *n=7* by a formula, analogical formula 3. Standard deviation from the mean measured volume was calculated by formula 4. The results of calculations are presented in a table 7.

**Table 7.** Mean values of height and volume of Staropetrovsky vessel layers between the marks; *m* - mark, *j* - number of the layer, $h_{l\_j\_m}$ - the average height of the layer *j*, $V_{l\_j\_m}$ - the average volume of the layer *j*, $\sigma_{l\_j}$ - the standard deviation of the average volume of the layer *j*.

| *m* | *j* | $h_{l\_j\_m}$, (cm) | $V_{l\_j\_m}$, (cm$^3$) | $\sigma_{l\_j}$, (cm$^3$) |
|---|---|---|---|---|
| 1 | 0 | 0.8 | 36.8 | ±0.5 |
| 2 | 1 | 1.6 | 103.4 | ±2.9 |
| 3 | 2 | 1.5 | 118.3 | ±5.7 |
| 4 | 3 | 1.3 | 155.8 | ±4.4 |
| 5 | 4 | 1.0 | 114.3 | ±4.8 |
| 6 | 5 | 0.9 | 156.3 | ±6.6 |
| 7 | 6 | 1.3 | 159.8 | ±3.8 |
| 8 | 7 | 1.2 | 145.2 | ±5.6 |
| T | 8 | 1.2 | 145.4 | ±5.7 |
| Edge of corolla | 9 | 0.6 | 45.9 | ±6.5 |

For verification of hypothesis about clepsydra we considered layers 1-7 between marks 1 and 8. The layer of water between a bottom and first mark was not examined by us, as his volume was considerably fewer volumes are between other marks. From our point of view, for the beginning of counting out of time, a vessel each time had to be filled by water to the first mark. It could be expedient from the point of view of control of integrity of bottom and control of horizontal or the same level of setting of vessel near to horizontal. In last case, as auxiliary, could be used, just, the same mark that marks the first on drawing located, approximately, on the same height, what the first mark of measuring.

On results measuring the total measured volume of Staropetrovsky vessel is equal $V_{m\_gen}$ =1181.1 cm$^3$ (таб. 3). The total measured volume of layers of vessel between extreme (1th and 8th) marks is equal $V_{l\_mark}$=953.1 cm$^3$. For comparison, the total calculated volume of vessel between analogical (1th and 8th) marks on drawing is equal $V_{mark}$=968.9 см3. I.e. the total calculated volume between these marks differs from measured volume less, than on 2%, that testifies about good quality of drawing of profile of vessel and adequacy of the chosen mathematical model for the calculation of it volume. Calculated by the formula, to the analogical formula 3, the mean measured volume of layer between nearby marks is equal $V_{l\_mean}$=136.2 cm$^3$. Calculated by the formula, to the analogical formula 4, corresponding standard deviation is equal $\sigma_{l\_mean}$=±21.7 cm$^3$. Thus, the measured volume of layer of vessel between two nearby marks appeared equal 136.2±21.7 cm$^3$.

The measured volumes of all layers between marks correspond to this range. An exception makes a sixth layer only. However it volume exceeds the high bound of the indicated range only on 1.9 cm$^3$ (1.2%). Such insignificant rejection can be attributed to the error of measuring, but, probably, it is related to the sloping location of seventh mark. Maybe, in measuring it was necessary to be oriented on it lowers edge.

Expected volume of layer of vessel between two nearby lateral marks, as be indicated higher, is equal 139.4±25.4 cm$^3$. It mean value is only insignificant, on 3.2 см$^3$ or 2.4%, exceeds the mean value of the measured volume of layer between nearby marks. This fact confirms possibility of realization of calculations of his volume with the use of mathematical model, being a complex of the truncated abrupts, approximating the profile of vessel, on drawing of vessel.

If we will consider overhead part of rim of "wheel" as a mark, then a general volume between extreme marks (by the first mark and mark T) will be equal ≈1098.4 cm$^3$, and the volume of layer between an eighth mark and mark T will be equal ≈145.4 cm$^3$. I.e. it corresponds to the mean value of the measured volume of layer between other nearby marks on a side. Thus, overhead part of rim of "wheel" fully could use, as another mark.

**The ancient water clock**

The best-known type of ancient vessels with vertical marks is water clock (clepsydra).

The earliest mention of a water clock was found in the texts of cuneiform tablets of collections Enuma - Anu - Enlil (XVII-XII century BC) and MUL.APIN (VII century BC) [38]. In these tablets speaks about water clock in connection with the board to the guards in the daytime and nighttime.

However, the ancient water clocks and fragments thereof from Mesopotamia were not found until now. Is therefore their appearance remains unclear. It is assumed that in Babylon could be used water clocks of cylindrical shape [39].

The oldest of discovered water clocks were found in Egyptian territory at Karnak and date the era of Amenhotep III (XIV c. BC) (Fig. 4 a). In Egypt, was also found an inscription with description the water clock in the tomb of Amenemhet (XVI c. BC), dignitary of times of Amenhotep I, which asserted that Amenemhet was their inventor [40].

Fragments of the Egyptian clepsydra and reduced copies were found too. They dated Hellenistic and Roman periods [41 - 46].

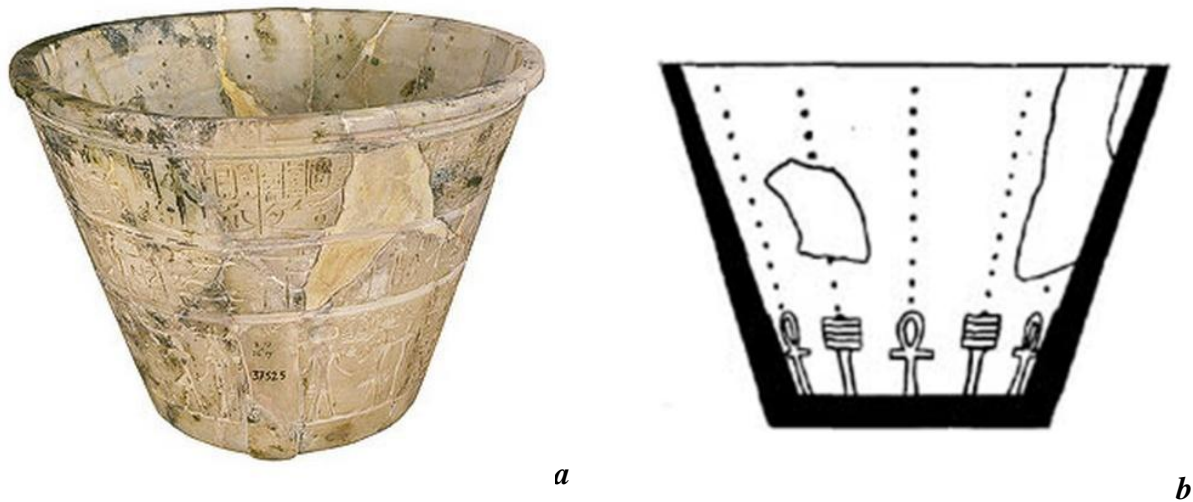

*a*  *b*

**Figure 4.** Karnak clepsydra: *a* - photo of clepsydra (Cairo, Egyptian Museum)[5], *b* - scheme of Karnak clepsydra scales on its inner surface [47].

The ancient water clocks could have a different shape (inverted truncated cone, prism, cylinder, etc.) And they measured the volume of flowing or flowing water. Water clock could be filled and flowing down type [48].

Water clocks were used to measure time indoors, often in the temples in the commission of divine service when required to consider the time. Every hour of the day was dedicated to a deity, and a special prayer dedicated to each deity. Water clocks were an important tool for determining the time in astronomical observations at night [49].

Counting out of time in the Egyptian clepsydras was produced on the graduated scales inflicted on the internal surface of vessel. Clepsydra from Karnak had 12 similar scales of different length, probably, on one on every month of year (fig. of 4b). They were divided into 11 intervals, probably, allowing to measure time from the end of 1th hour to the end of 12th o'clock of night. By an initial point for measuring, maybe, rising or culmination of certain star served, whereupon counting out of clock was produced regardless of star supervisions [50]

In some other standards of clepsydras also there were 12 scales, but divided already into 12 intervals. In a number of clepsydras distribution of scales on months was uneven [51]. Thus, there were not strict rules for marking of the Egyptian clepsydras, and they fully could change in different epochs.

---

[5]http://www.eternalegypt.org/EternalEgyptWebsiteWeb/HomeServlet?ee_website_action_key=action.display.text.viewer&language_id=1&element_id=60513&name=19741&image_name=http://www.eternalegypt.org/images/elements/19741_310x310.jpg&mode=1n&title=Water%20Clock

**Interpretation of Staropetrovsky vessel as clepsydra**

Eight the lateral marks are on Staropetrovsky vessel (ninth mark turned out to accidental scratches - see. above). They marked seven layers of the water. Each layer has volume 136.2±21.7 cm$^3$.

If to confront every layer between marks with a temporal interval equal to one hour, then by means of seven layers it was possible to measure seven o'clock of equal duration. If to take into account possibility of the use of overhead part of rim of "wheel" as another mark, then by means of such marking it was possible will measure eight hours. By means of such marking it was possible to measure every hour with mean exactness $\Delta t_{l\_mean} = \sigma_{l\_mean} / V_{l\_mean} = \pm 0.16$ hour $= \pm 9.6$ minutes.

If to assume that corresponds every layer of Staropetrovsky vessel one hour duration 60 minutes or 3600 seconds, then to the interval to time equal to one second, the volume of water of $V_{1sec} \approx V_{l\_mean}/3600 = 0.04$ cm$^3$. It is known that the volume of ordinary water drop averages[6]: 0.03-0.05 cm$^3$. Long time, up to XIX of century, a "drop" was the minimum unit of chemist measure[7] [52]. I.e., if with a help of Staropetrovsky vessel measured time, and corresponded every his mark one hour equal 1/24 parts of sunny twenty-four hours, then water in a vessel had to act at a speed of equal, approximately, to one drop in a second. However, about a second, as unit of time, for certain it is known only from 1000 AD [53], although absence of reliable certificates does not yet talk about impossibility of existence of this unit in more early epochs. Existence in more early periods and self origin of second fully could be related to watching the pulse of the grown healthy man in the quiet state, that, as a rule, is equal to 60 shots in a minute or one blow in a second[8].

Thus, equality of volume of water, corresponding to one second (to the least unit of time), volume of one water drop (to the least ancient unit of volume of liquid), also can be examined, as a certificate in behalf on using of Staropetrovsky vessel as clepsydras of story type.

In ancient Babylon weight of water in clepsydras was measured in mina (mana) [54]. Coming from duration of twenty-four hours equal 6 mina [55-57] and weight one mina, being, approximately, in a range from a 460 g a to 540 g [58], it is possible to define that weight of water for measuring of one hour was in a range from a 115 g a to 135 g, and weight corresponded one second in the range from 0.03 g to 0.04 g.

Thus, taking to account that the average density of fresh water ≈1 g/cm$^3$ at temperatures from +15º C to +20º C [59], that a volume of fresh water, corresponding one mina, is in a range from 460 cm$^3$ to 540 cm$^3$ (consequently, on the average a volume one mina is equal ≈500 cm$^3$), volume of water, necessary for measuring of 1 hour is in a range from 115 cm$^3$ to 135 cm$^3$, and for measuring of one second - from 0.03 cm$^3$ to 0.04 cm$^3$. The volume of water for measuring of one hour by means of Staropetrovsky vessel exceeds a high bound for traditional mina, approximately, on 1.2 cm$^3$ (0.9%). So small size of difference it is fully possible to ignore, attributing her to the error of measuring.

In Old Babylonian astronomic texts - in the tablets of BM 17175 + 17284 (published in the application to edition of MUL.APIN [60]) – see instructions for determining the duration of day and night - day and night watchmens - depending on the seasons. In days an equinox for measuring of duration of day required 3 mina waters, and nights - 3 mina. In the day of summer solstice for a day - 4 mina, and nights - 2 mina. In the day of winter solstice for a day - 2 mina, and nights - 4 mina.

---

[6] http://ru.wikipedia.org/wiki/Капля
[7] http://en.wikipedia.org/wiki/Minim_(unit)
[8] http://www.nlm.nih.gov/medlineplus/ency/article/003399.htm

This text served a prototype for later texts in "Astrolabes", in the table of XIV of series of EAE and in MUL.APIN [61].

An interesting fact is that indicated in the table of BM 17175 relation of duration of day in the day of summer solstice to duration of day in an equinox and to duration of day in winter solstice - 4:3:2 (like for night, but upside-down), most exactly corresponds to the not latitudes of Mesopotamia (approximately, 30º - 38º N), but to the latitudes between 45º N and 50º N (таб. 8, 9). In a table 8 results over of calculations of duration of day (from rising to setting of Sun) are brought in minutes for 1500 B.C. for longitude equal 44º $E$ (the choice of longitude of fundamental value does not have, as duration of day and night depends only on a latitude) and for latitudes in a range from 30º N to 50º N. Calculations were produced by us by means of the astronomic program RedShift - 7 Advanced. In a table 9 the results of count of duration of day are presented in mina for the same latitudes.

To correspond to the correlation indicated in the table of BM 17175, at duration of day in an equinox equal 725 minutes, duration of day in summer solstice must be equal 967 minutes, and in winter solstice 483 minutes. Thus, duration of day less than, than it is indicated in the table of BM 17175 in summer solstice, for 30º N approximately on 2 hours, for 35º N on 1.6 hour, for 40º N on 1.1 hour, for 45º N on 0.5 hour, and for 50º N - a more hour is on 0.3. Duration of day more than it is indicated in the table of BM 17175 in winter solstice, for 30º N approximately on 2 hours, for 35º N on 1.7 hour, for 40º N on 1.2 hour, for 45º N on 0.6 hour, and for 50º N - a less than hour is on 0.1. Correlations of duration of day and, accordingly, nights will be exactly such, as indicated in a table, in a range from 47º55′ N to 49º25′ N with exactness 5′.

**Table 8.** Duration of day in minutes on different latitudes, expected for 1500 BC; $t_{day}$ - duration of daily time in minutes.

|  | $t_{day}$, (minute) | | | | | |
|---|---|---|---|---|---|---|
|  | 30º $N$ | 35º $N$ | 40 º $N$ | 45º $N$ | 48º $N$ | 50º $N$ |
| summer solstice | 845 | 871 | 901 | 939 | 965 | 983 |
| equinox | 725 | 725 | 725 | 725 | 725 | 725 |
| winter solstice | 608 | 582 | 553 | 518 | 493 | 475 |

**Table 9.** Duration of day in mina on different latitudes, expected for 1500 BC; $t_{mina}$ - duration of daily time in mina.

|  | $t_{day}$, (mina) | | | | | |
|---|---|---|---|---|---|---|
|  | 30º $N$ | 35º $N$ | 40 º $N$ | 45º $N$ | 48º $N$ | 50º $N$ |
| summer solstice | 3.5 | 3.6 | 3.7 | 3.9 | 4.0 | 4.1 |
| equinox | 3.0 | 3.0 | 3.0 | 3.0 | 3.0 | 3.0 |
| winter solstice | 2.5 | 2.4 | 2.3 | 2.1 | 2.0 | 2.0 |

Staropetrovsky vessel has a volume between the extreme marks (taking into account the mark of T) ≈of 1098.4 cm$^3$=2.03 mina≈2 mina (where 1 mina=540 cm$^3$). Thus by means of nine marks (eight on an internal side and mark T on external) of Staropetrovsky vessel it was possible to measure eight o'clock of equal duration like the Babylonian clock, at filling with of vessel water with the same speed, as well as in the Babylonian clepsydras, is one drop in a second.

Approximately eight hours a day lasts in winter solstice (night in summer solstice) on the latitude of finding out Staropetrovsky vessel. For measuring of this interval of time a vessel had to be filled one time. In the day of summer solstice, or in night of winter solstice, Staropetrovsky vessel had to be filled twice.

And during whole twenty-four hours - three times. I.e. by means of Staropetrovsky vessel it was easily to produce measuring of duration of day and night in summer and winter solstices by rule indicated in the table of BM 17175. And, if to take into account that the latitude of finding out Staropetrovsky vessel is equal to $48^0 13'$ $N$ and gets in the range of latitudes optimally corresponding to the rule from this table, then it is possible to suppose that, at least, the mediated contacts between the frame population of North Black Sea region and population of Ancient Mesopotamia in Late Bronze Age existed nevertheless.

Ancient water clock of Mesopotamia have not found until now. And no one knows exactly how they looked. It can be assumed that they could be similar to Staropetrovsky vessel. Externally it is a simple hand-made vessel with inconspicuous nail depressions on the inner side surface. The analogy between the Mesopotamian water clock and Staropetrovsky vessel may further help researchers in search of as fragments or whole ancient Mesopotamian water clocks. Known ancient Egyptian water clock - Karnak clepsydra - greatly exceeding the volume of Staropetrovsky vessel (Fig. 4). Thus, the working volume of Karnak clepsydra is approximately 22 liters. It is believed that it was intended to measure 12 hours at a rate of 10 drops per second, with the volume of droplets 0.05 cm$^3$ [62, 63].

I.e. for measuring of one hour the 1.8 liters was required, approximately. It more than in three times exceeds a volume, necessary for the analogical measuring both in the Mesopotamia clepsydras and in a Staropetrovsky vessel. And, speed of filling of such clepsydra must almost tenfold exceed speed of filling of the Mesopotamia clepsydras.

However, it is known that in Ancient Egypt there was unit of volume *hinu* or *hin (jar)* equal 480 cm$^3$ [64].

Weight of fresh water of such volume is approximately equal on weight one Mesopotamia *mina*. Maybe, *hinu* and *mina* have a homogeny, and *hinu* was used in clepsydras like *mina* in more simple variants of clepsydras, than Karnak clepsydra. So, in writing of the Egyptian name of unit of volume *hinu* (fig. 5a) and in writing of hour (fig. 5b), unlike writing of other units of volume and time, there is hieroglyph, interpreted as a vessel - Nw or Nu jar/pot (fig. of 5c).

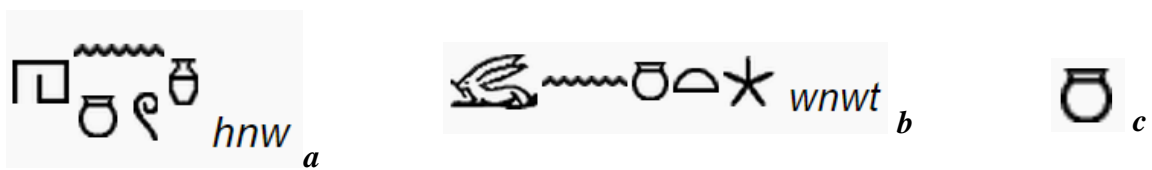

**Picture 5.** Egyptian inscriptions, containing the hieroglyph Nu pot: ***a*** - unit of volume *hinu*[9], ***b*** - hour[10], ***c*** – Nu pot[11].

Is believed that the Nu pots often depicted on the frescoes and sculptures of pharaohs, bringing in them sacrificial gifts, presumably, wine or fragrant oils (Fig. 6).

---

[9] http://en.wikipedia.org/wiki/Ancient_Egyptian_units_of_measurement#cite_note-CR-2

[10] http://en.wikipedia.org/wiki/Ancient_Egyptian_units_of_measurement#cite_note-CR-2

[11] http://hieroglyphes.pagesperso-orange.fr/Index%20W-b.html#W10    ;
http://en.wikipedia.org/wiki/List_of_portraiture_offerings_with_Ancient_Egyptian_hieroglyphs

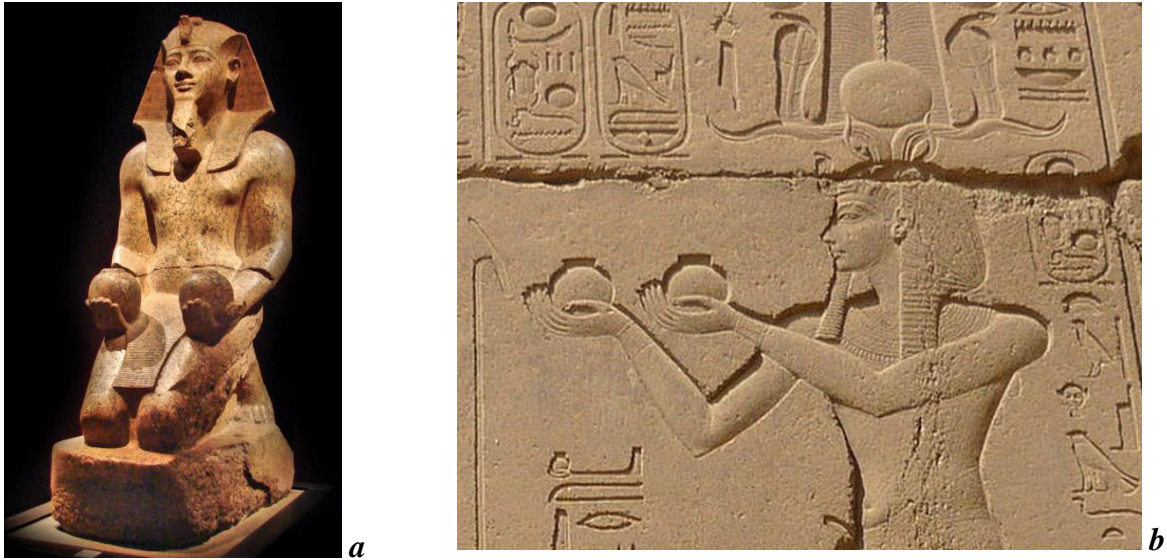

**Figure 6.** Images of Nu pot: *a* - the statue of pharaoh Amenhotep II (XV-XIV centuries BC), *b* - a fragment of the wall relief with the image of pharaoh Horemheb (XIV-XIII centuries BC).

Dimensions of spherical parts of the vessel are shown in all images about the same and are comparable with the palm of the human hand. Therefore, knowledge of the volume of even a single vessel could allow approximately estimate a traditional volume for the entire series of vessels. Well-preserved vessel Nu from Egyptian alabaster (onyx marble), which belonged to pharaoh Unis (XXIV century BC), kept in the Louvre (Fig. 7).

Published exact dimensions of the vessel: height is 17 cm, maximum diameter is 13.2 cm [65]. At a thickness of the walls ≈ 0.3 cm determined by photo (see the edge of the corolla), the internal volume of the spherical part of the vessel, calculated by the formula sphere volume, equal ≈1047 $cm^3$. This volume is approximately equivalent to 2 hinu (960 $cm^3$).

The difference is just 9%. Volume of Staropetrovsky vessel between the extreme markers equal 1098.4 $cm^3$, which is only 5% greater than the volume of the spherical Nu pot from the Louvre. Thus, the volume of water contained in the spherical part of the Nu pot corresponds to the volume of water needed to measure eight hours of time in Mesopotamian tradition (and in the case of Staropetrovsky vessel).

God Horus as a falcon, who holds in his paws two characters Shen - symbols of longevity and eternity, is depicted on the vessel from the Louvre. The basic idea of the totality of the image on the surface of the vessel is treated as "eternal renewal of life." Such an idea is associated with the notion of finite time of life and its constant change - the movement that may also testify in favor of the version about the use of Nu pots for measuring time.

Among Egyptian hieroglyphics, containing images of Nu pot, there is hieroglyph W123 (Fig. 8). On it the Nu pot crossed by horizontal lines, dividing it into layers. We believe that this character may reflect the tradition of using Nu pot as a metric vessel, including for measuring the volume of water in the water clock.

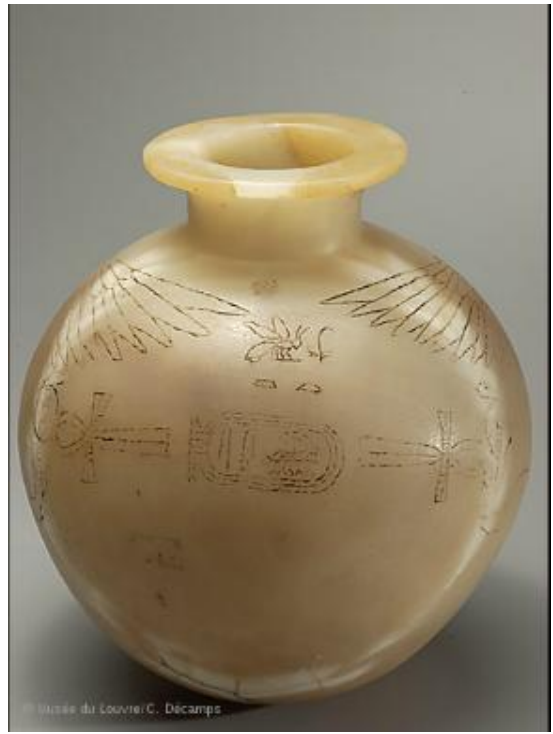

**Figure 7.** Photo of the vessel Nu from the Louvre[12].

On the hieroglyph W124 represented at once three Nu pots. Namely three vessels of water, with a volume equivalent to the volume of the Nu pot, it is necessary to measure the duration of the 24 hour day in the Mesopotamian tradition (and in the case of Staropetrovsky vessel). Hieroglyphics W25 and W101 depict Nu pots with legs. We believe that in these hieroglyphs can be reflected motion time and/or water measured by the Nu pots in the case of using them as a water clock.

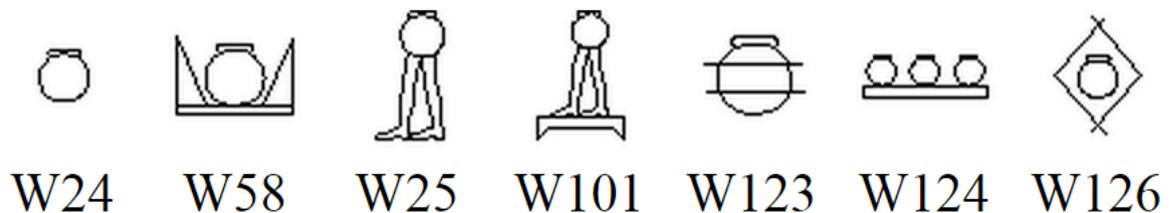

**Figure 8.** Egyptian hieroglyphs containing the image of the Nu pot[13]. Presented numbering of hieroglyphs based on the list of A.H. Gardiner [66].

The fragment of text is known also on the surface of fragment of the ruler or L- shaped sundial: "Hour in an elbow[14]. Jar from a copper, filled by water." [67]. Jar from a copper in this fragment researchers interpret, as clepsydra [68]. It is possible that copper jar from inscription on the ruler and vessel designated by a hieroglyph "Nu pot", - are the family types of vessels, being clepsydras of story type, the analogue of that can be Staropetrovsky vessel.

Clepsydras as a jar look like a Nu pot were maybe used in Ancient Egypt on the early stages of development of technology of measuring of time and were used in future for the domestic measuring, as were simple in making and did not require expensive materials for making. If it really so, and Nu pots, applied as clepsydras, could outwardly look, as the ordinary pots with marks inwardly, then now, knowing their basic external signs, could be purposefully try to find them among those already found ancient Egyptian pottery vessels and fragments thereof.

---

[12] http://www.louvre.fr/en/oeuvre-notices/vase-name-king-wenis
[13] http://hieroglyphes.pagesperso-orange.fr/Index%20W-b.html#W10
[14] elbow (cubit) is length unit

**Conclusion**

Thus, during the study was created, the geometric model of Staropetrovsky vessel with the marks on inner surface by a previously published archaeological drawing. With the help of statistical methods produce results that evidence in favor interpretation vessel as a water clock. The assumption that Staropetrovsky vessel is an ancient water clocks originated in the study of found in the Northern Black Sea coast oldest analemmatic sundial, belonging to the same epoch and archaeological culture [69, 70]. For marking sundials was necessary standard of time, which could be implemented with the help of water clock similar with Staropetrovsky vessel. In order to confirm hypothesis about the water clock a series of measurements of the volume of Staropetrovsky vessel before each of the marks on the lateral surface. Comparison and analysis of the results of calculations and measurements of Staropetrovsky vessel allowed to be considered the hypothesis about water clock practically proven.

Staropetrovsky vessel, how water clock, could be used to measure time indoors during the rituals associated with a particular time of day and to determine the time during the night hours for astronomical observations.

The earliest evidence of the existence of a water clock in territory of Europe is a mention of the use clepsydra by Empedocles in Ancient Greece (island of Sicily) in V centuries BC [71, 72]. Since Staropretrovsky vessel from the Central Donbass is dated XV-XIV centuries BC, it can be argued that at the moment it is the most ancient water clocks found on the territory of Europe.

**Acknowledgement**